# Dual Fronts Propagating
# into an Unstable State

*F.J. Elmer[1], J.-P. Eckmann[2,3], and G. Hartsleben[2]*

[1]Institut für Physik, Universität Basel, CH-4056 Basel, Switzerland

[2]Dépt. de Physique Théorique, Université de Genève, CH-1211 Genève 4, Switzerland

[3]Section de Mathématiques, Université de Genève, CH-1211 Genève 4, Switzerland

**Abstract.** The interface between an unstable state and a stable state usually develops a single confined front travelling with constant velocity into the unstable state. Recently, the splitting of such an interface into *two* fronts propagating with *different* velocities was observed numerically in a magnetic system. The intermediate state is unstable and grows linearly in time. We first establish rigorously the existence of this phenomenon, called "dual front," for a class of structurally unstable one-component models. Then we use this insight to explain dual fronts for a generic two-component reaction-diffusion system, and for the magnetic system.

## 1. Introduction

Parabolic differential equations, such as the (complex) Ginzburg-Landau equation [CE, CH], occur in many domains of physics. We are here interested in this equation when considered on an infinite domain. Then, a well-known phenomenon is the formation of fronts (travelling waves), which are solutions whose shape is fixed in a frame moving with speed $c$. For example, for the equation

$$\dot{u} = u'' + u - u^3 ,\tag{1.1}$$

one has the explicit solution

$$u(x, t) = \frac{1}{2}\left(1 - \tanh(2^{-3/2}(x - 3 \cdot 2^{-1/2}t))\right) ,$$

which moves with speed $c = 3 \cdot 2^{-1/2}$. In a recent paper, a new phenomenon was observed [EBS]: A model for the dynamics of an easy-plane ferromagnetic material was considered, and it led to the surprising discovery that several concurrent front speeds can be observed, with a leading front connecting an unstable state to an intermediate unstable state, which in turn gives rise to a transition to a stable final state. As time goes on, the spatial region in which the intermediate state is present is seen to *grow*. We shall call this phenomenon a *dual front*. The aim of this paper is to explain this somewhat counterintuitive fact, and to put it into a more general perspective. Diffusion problems with several coexisting fronts have been discussed also elsewhere in the literature, see e.g., [PNJ], [GPP]. The mechanism described in those papers is, however, of a different nature from the one studied here, since the intermediate state is *stable*. Before starting the general discussion of dual fronts, we recall the model of [EBS].

In that paper, a one-dimensional ferromagnetic material is considered, which is described by the magnetization $M(x) \in \mathbf{R}^3$, with $|M(x)| = 1$. Supposing that the material has an easy-plane orthogonal to the direction $\zeta$, one obtains for the free energy

$$\begin{aligned}W &= \frac{1}{2}\int dx\ \left(|M'|^2 + M_\zeta^2\right)\\&= \frac{1}{2}\int dx\ \left(\vartheta'^2 + \varphi'^2 \sin^2\vartheta + \cos^2\vartheta\right) ,\end{aligned}\tag{1.2}$$

where $'$ stands for derivative with respect to $x$, $M_\zeta$ is the component of $M$ in the direction $\zeta$, and $\varphi$ and $\vartheta$ are spherical coordinates with respect to polar direction $\zeta$. It is assumed that the dynamics is driven by the Landau-Lifschitz equation [HC], i.e.,

$$\dot{M} = -\gamma(M \times H) - \lambda M \times (M \times H) ,$$

where $H = -\delta W/\delta M$, and that damping is sufficiently high such that the torque part $-\gamma(M \times H)$ can be neglected. Setting $\lambda = 1$ yields the evolution equation

$$\begin{aligned}\dot{\vartheta} &= \vartheta'' + (1 - \varphi'^2)\sin\vartheta\cos\vartheta ,\\\dot{\varphi} &= \varphi'' + 2\varphi'\vartheta'\cot\vartheta .\end{aligned}\tag{1.3}$$

In terms of the local wave number $k = \varphi'$, the second equation in (1.3) becomes

$$\dot{k} = k'' + 2(k(\ln \sin \vartheta)')' \ . \tag{1.4}$$

We emphasize that the singularities at $\vartheta = n\pi, n \in \mathbf{Z}$ are artificial, as can be seen by considering the complex variable $A(\vartheta, \varphi) = \sin \vartheta e^{i\varphi}$. Equation (1.3) is equivalent to

$$\dot{A} = A'' + \left( 1 - |A|^2 + |A'|^2 + \frac{((|A|^2)')^2}{4(1 - |A|^2)} \right) A \ , \tag{1.5}$$

which is regular around $A = 0$.

The following uniform stationary states of (1.3) will be considered:

$$
\begin{aligned}
1) \quad & \vartheta(x) = \pi/2, \ \varphi(x) = k_0 \cdot x \ , \\
2) \quad & \vartheta(x) = 0 \ .
\end{aligned}
$$

Solutions of the first type are called by [EBS] *spiral states*, we call the second one the *zero-solution*. Linear stability analysis shows that spiral solutions are stable for $|k_0| < 1$ and have one unstable direction pointing in the $\vartheta$-direction when $|k_0| > 1$. The solution $\vartheta(x) = 0$ is unstable.

If we choose an initial condition that connects a stable state to an unstable spiral state, we expect that a front will invade the unstable spiral. In [EBS] this phenomenon was studied in detail, for typical initial conditions connecting the stable state with $k_0 = 0$ to an unstable spiral state, i.e., for

$$\lim_{x \to \pm\infty} \vartheta(x, t = 0) = \frac{\pi}{2} \ , \quad \lim_{x \to \infty} \varphi'(x, t = 0) = k_0 > 1 \ , \quad \text{and} \quad \lim_{x \to -\infty} \varphi'(x, t = 0) = 0 \ .$$

It was observed that for $k_0 > \sqrt{2}$ the solution converges towards a leading front that connects the unstable spiral solution to the zero-solution, which in turn gives rise to a transition to the stable solution with $\varphi'(x) = 0$. The size of the region in which $A$ is close to zero was seen to *grow* with time; see Fig. 1. This phenomenon of a growing intermediate region is the main object of this paper.

Furthermore, a *third* front, which we call "ghost front," can be seen in the simulations. It is the line going through zeros of the amplitude $|A|$, which is visible in Fig. 2. Its origin will be explained in Chapter 5.

The structure of this paper is as follows. We will show that the phenomenon described in the magnetic model above is a special case of a more general set of similar phenomena which can be observed in many parabolic models. After a discussion of fronts in Section 2, we will start by exhibiting this phenomenon in 1-component systems in Section 3, then we describe a structurally stable situation with 2 components in Section 4, and finally describe the magnetic model in Section 5. Most of our results cannot be rigorously proven, but a few facts can be established rigorously. Some of them are collected in the Appendix.

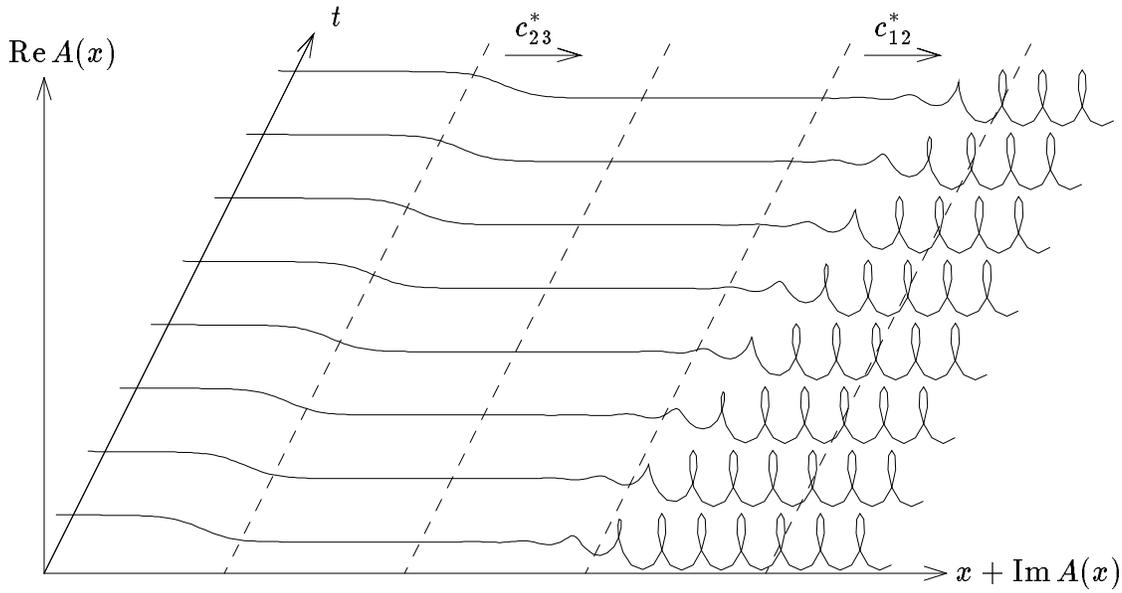

**Fig. 1**: Several time-frames ($t = t_0$ to $t = t_0 + 3$) for the evolution of Eq.(1.3), with $k_0 = 2$. One can clearly see the leading front moving with a speed $c_{12}^*$ and the trailing front moving with different speed $c_{23}^* < c_{12}^*$.

$$\frac{|A|}{2} + \left(1.5 - \log \frac{|A|}{|A| + 5 \cdot 10^{-9}}\right)^{-1}$$

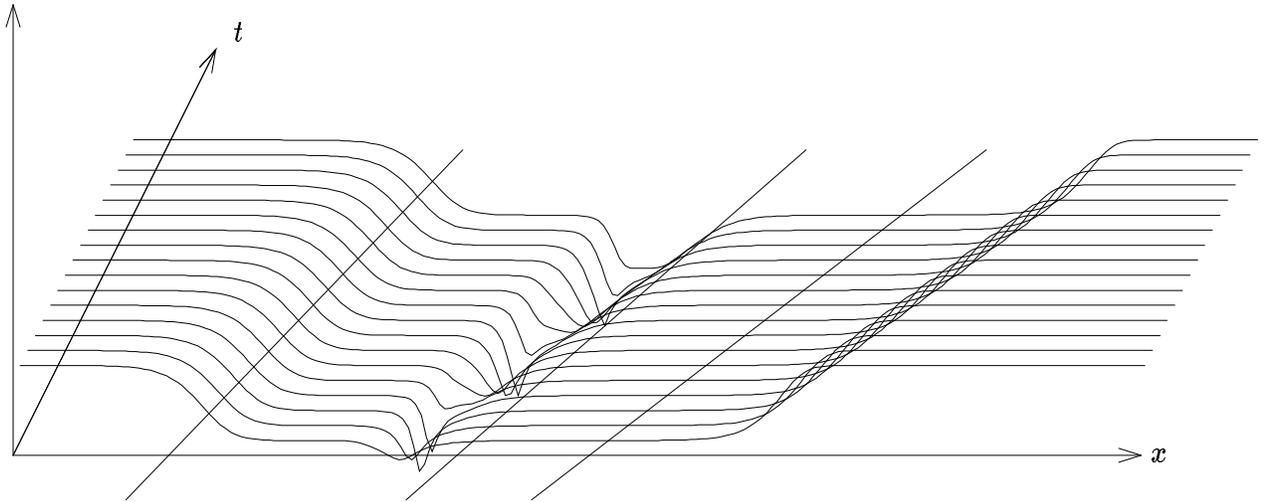

**Fig. 2**: Several time-frames ($t = t_0$ to $t = t_0 + 3$) for the evolution of Eq.(1.3), with $k_0 = 2$. One can clearly see the leading front moving with speed $c_{12}^*$ and the trailing front moving with speed $c_{23}^*$. In the center, one recognizes the ghost front, moving with speed $c_{\text{ghost}}$. The vertical scale has been chosen in such a way to make the phase-slip points in the ghost front better visible (these are very small amplitudes). The three solid lines indicate the speeds. The middle line is fitted through the minima of the amplitudes. The boundaries of the "central valley" expand with the speeds $c_{12}^*$ and $c_{23}^*$, since the overall amplitudes are going to zero with the same speeds as the trailing parts of the two fronts.

## 2. Fronts and their speeds

Before we can discuss the existence of dual fronts, we need a precise description of such notions as "critical speed" and "minimal speed" for fronts of parabolic equations. This discussion will follow mostly the original definitions of Aronson and Weinberger [AW]. Let us consider for simplicity an equation of the form

$$\dot{u} \; = \; u'' + f(u) \;, \tag{2.1}$$

where dots stand for temporal and primes for spatial derivatives, and where $u(x,t)$ is a real function. We assume that $f \in \mathcal{C}^1$, and that, on the interval $[\ell, r]$, the function $f$ satisfies $f(x) > 0$ for $x \in (\ell, r)$, and $f(\ell) = f(r) = 0$, see Fig. 3.

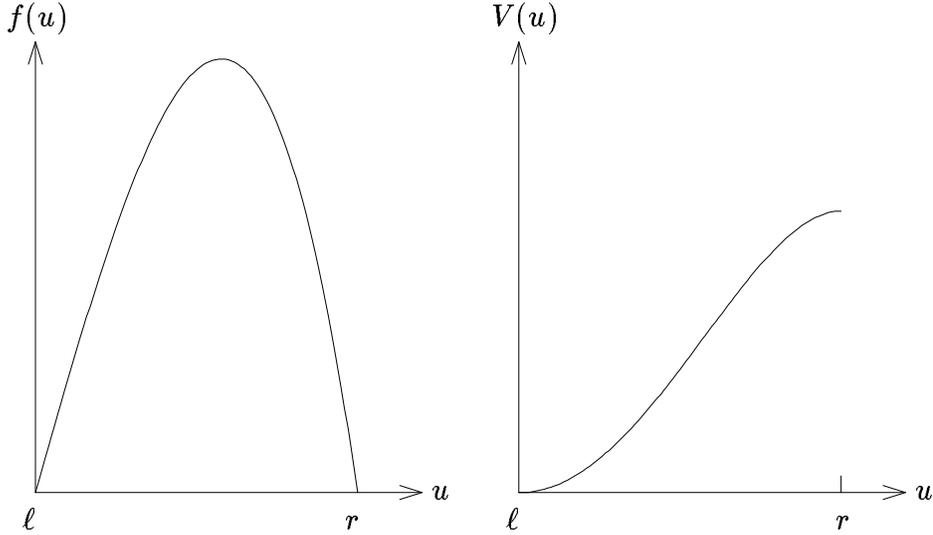

**Fig. 3**: Typical shape of the force $f$ and of the corresponding potential $V$.

As is well known [AW], the existence of fronts (travelling waves) and their speed $c$ is discussed by going to one (or several) moving frame(s) and studying solutions of the form $u(x,t) = q(x - ct)$. Then $q$ satisfies the "Hamiltonian" equations

$$q'' \; = \; -cq' - V'(q) \;, \quad V'(q) = f(q) \;, \tag{2.2}$$

or, equivalently,

$$\begin{aligned} q' &\; = \; p \;, \\ p' &\; = \; -cp - f(q) \;. \end{aligned} \tag{2.3}$$

One sees that $c$ appears as a friction, and $f$ as the derivative of a potential $V$ in which the particle is moving, so that the zeros of $f$ correspond to equilibria of this potential. We begin by discussing the *existence* of fronts, in the case of the potential $V$ with critical points at $\ell$ and $r$. (See also similar discussions in [CE], [AW], [PNJ].) Using the picture of a Hamiltonian flow with friction $c$, we observe that when $c = 0$, a particle released at $r$ will "fall down" and traverse the critical point $\ell$ ("overshooting").

**Definition.** The *minimal speed* associated with $f$ is the smallest friction $c$ for which the particle starting at the point $r$ with zero velocity reaches the point $\ell$ without overshooting. We call this speed $c_{\ell r}^{\min}$.

**Remark.** In the case under discussion, the *existence* of $c_{\ell r}^{\min}$ is obvious. However, in the next section we will see an example where certain minimal speeds do not exist.

**Definition.** The *critical speed* $c_{\ell r}^*$ is the infimum of those frictions for which there is a initial velocity at $r$ (which may be zero or *non-zero*) allowing a particle leaving $r$ to the left to reach $\ell$ without overshooting.

**Remark.** The critical speed always exists. See Appendix A for a precise definition of $c_{\ell r}^*$. We also state there that $c_{\ell r}^{\min} = c_{\ell r}^*$ for the case considered in this section.

An important issue in the theory of fronts is their "selection," i.e., the question which of several possible speeds will be chosen by an arbitrary initial condition. One has the following fundamental theorem about front *selection* that was proven in [AW]: it shows that the dynamics chooses the front with speed $c_{\ell r}^*$. Assume that $f$ is as above.

**Lemma 2.1.** [AW], [CE]. *Assume that the initial condition $u(x, 0)$ takes values in $[\ell, r]$, and reaches the limits*

$$\lim_{x \to \infty} u(x, 0) = \ell \,,$$

$$\lim_{x \to -\infty} u(x, 0) = r \,,$$

*for finite $x$. Then, one has*

$$\lim_{t \to \infty} u(x + ct, t) = \begin{cases} \ell \,, & \text{if } c > c_{\ell r}^* \,, \\ r \,, & \text{if } c < c_{\ell r}^* \,. \end{cases} \tag{2.4}$$

Thus, if we move slower than the critical speed, we are "behind" the front and will see eventually the state $r$, and if we move faster, we see $\ell$.

## 3. A one-component example

We now discuss a one-component model which exhibits a dual front. This model is again given by equations of the form

$$\dot{u} = u'' + f(u) \,, \tag{3.1}$$

where we now suppose that $f$ is a non-negative $\mathcal{C}^1$ function, $f(u) \geq 0$ in $[0, 1]$ with *three* zeros at $S_1 = 0$, $S_2 = \frac{1}{2}$, and $S_3 = 1$; cf. Fig. 4. The choice of $S_2 = \frac{1}{2}$ is inessential.

Coming back to the discussion of critical and minimal speeds, we can consider the three intervals $(\ell, r) = (S_1, S_2), (S_1, S_3)$, and $(S_2, S_3)$. In the intervals "12" and "23", the discussion of the previous section applies without change: The minimal speeds exist, and are equal to the critical speeds, which we denote $c_{12}^*$ and $c_{23}^*$, respectively. The critical speed $c_{13}^*$ for fronts connecting $S_1$ and $S_3$ exists by definition, but there is not necessarily an orbit with minimal friction connecting $S_3$ with $S_1$, since the potential may be such that any friction which lets the particle traverse $S_2$ must be so weak that the particle will necessarily overshoot at $S_1$. In fact, this is the case for the potential of Fig. 4. On the other hand, and this is the main point of our choice of $V$, non-existence of $c_{13}^{\min}$ is equivalent to the inequality $c_{12}^* \geq c_{23}^*$. As we will see, the strict inequality, $c_{12}^* > c_{23}^*$, implies the phenomenon of a dual front.

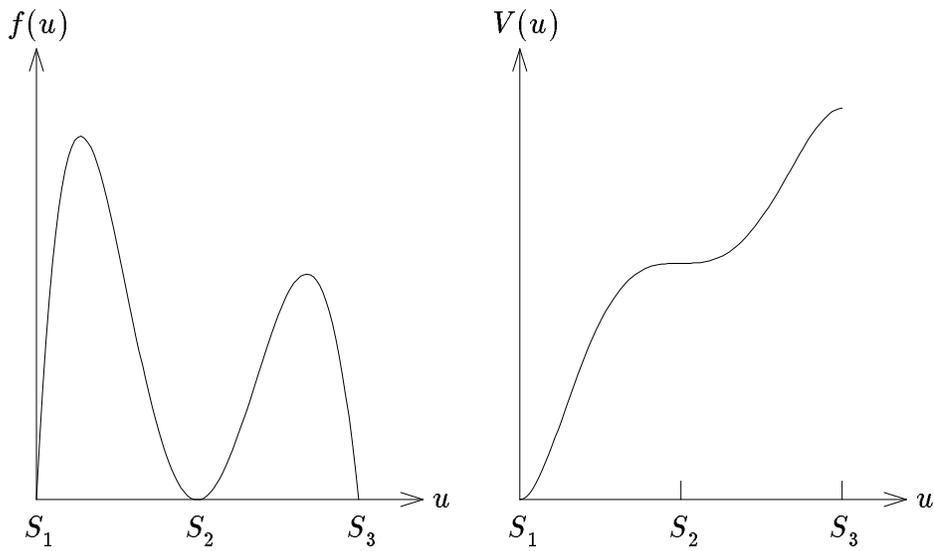

**Fig. 4**: Typical shape of the function $f$ and of the potential $V$.

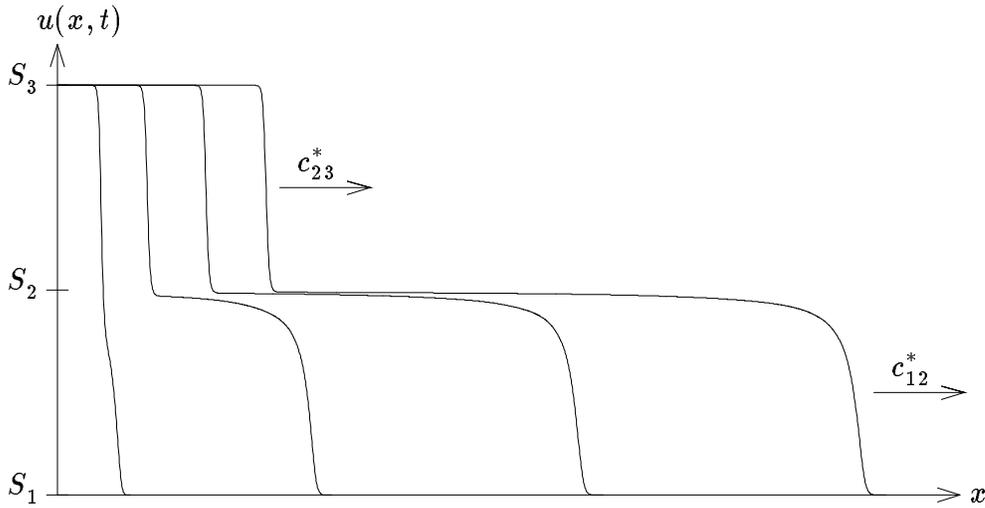

**Fig. 5**: The intermediate region grows in size as time advances. The frames shown are at $t = 5$, $t = 50$, $t = 100$, and $t = 150$ for the Eq. (3.1).

We apply the Lemma 2.1 to the two subintervals "12" and "23". This will prove that under the conditions mentioned above, the size of the intermediate state grows linearly in time for all initial conditions that approach the limiting values sufficiently fast. We have a dual front:

**Proposition 3.1.** *Let $f$ be such that $c_{12}^* > c_{23}^*$. Then, for any initial condition taking values in*

$[S_1, S_3]$ *and reaching the limits*

$$\lim_{x \to \infty} u(x, 0) = S_1 \,,$$

$$\lim_{x \to -\infty} u(x, 0) = S_3 \,,$$

*for finite* $x$, *an observer moving with speed* $c$ *will see the limit*

$$\lim_{t \to \infty} u(x + ct, t) = \begin{cases} S_3 \,, & \text{if } c_{23}^* > c \,, \\ S_2 \,, & \text{if } c_{12}^* > c > c_{23}^* \,, \\ S_1 \,, & \text{if } c > c_{12}^* \,. \end{cases}$$

**Remark.** The result above can be interpreted as follows: If we look in the laboratory frame at a position $ct$, with $c \in (c_{23}^*, c_{12}^*)$, we will find $u$ very close to $S_2$ if $t$ is sufficiently large. Thus, the state $S_2$ is visible on an interval whose length is of order $(c_{12}^* - c_{23}^*)t$. On the other hand, at $x = ct$, with $c > c_{12}^*$, $u$ will be close to $S_1$. At $x = ct$ with $c < c_{23}^*$, we see the state $S_3$.

**Proof. Case $c_{12}^* > c > c_{23}^*$:**
Consider the functions $\underline{\psi}(x, 0) = \min(u(x, 0), S_2)$ and $\overline{\psi}(x, 0) = \max(u(x, 0), S_2)$ as initial data for the Eq.(3.1). The Maximum Principle [F] tells us that

$$\underline{\psi}(x, t) \leq u(x, t) \leq \overline{\psi}(x, t) \ \text{ for all } (x, t) \in \mathbf{R} \times \mathbf{R}_0^+ \,.$$

Moreover, Lemma 2.1 tells us that $\lim_{t \to \infty} \overline{\psi}(x + ct, t) = S_2$ for all $c > c_{23}^*$ and that $\lim_{t \to \infty} \underline{\psi}(x + ct, t) = S_2$ for all $c < c_{12}^*$, which completes the proof in this case.

**Case $c > c_{12}^*$:**
The proof for the case $c > c_{12}^*$, is slightly more complicated. We consider a sequence of functions $f_\delta$ converging to $f$, as $\delta \downarrow 0$, such that $f_\delta(x) > 0$ for all $x \in (S_1, S_3)$. We choose them such that $f(x) + \delta \geq f_\delta(x) \geq f(x)$ and $f_\delta - f$ has support in $[S_2 - \delta, S_2 + \delta]$. Thus, the corresponding potential $V_\delta$ has only critical points at $S_1$ and $S_3$ and Lemma 2.1 can be applied to the interval $[S_1, S_3]$. The critical speed $c_{13}^*(\delta)$ is equal to the minimal speed $c_{13}^{\min}(\delta)$. The proof for the case $c > c_{12}^*$ will follow from

$$\lim_{\delta \to 0} c_{13}^*(\delta) = c_{12}^* \,. \tag{3.2}$$

To show Eq.(3.2), we first observe that $f_\delta(x) \geq f(x)$ implies $c_{13}^*(\delta) \geq c_{12}^*$, by monotonicity, cf. also Appendix A. On the other hand, for every $c > c_{12}^*$, there is a sufficiently small perturbation $f_\delta$ of $f$ for which a particle released at $S_3$ with zero velocity will actually arrive at $S_1$ without overshooting. To see this, note that for any $c > c_{12}^*$, there exists a $\mu(\delta) > 0$ for which a particle starting at $q = S_2 - \delta$, $0 \geq p \geq -\mu(\delta)$ will reach $S_1$ without overshooting. (This follows from the definition of $c_{12}^*$ as the limit of such initial data.) By continuity, there is a $\mu_0 > 0$ such that $\mu(\delta) > \mu_0$ for sufficiently small $\delta > 0$. On the other hand, if $\delta > 0$ is chosen sufficiently small, the orbit in the potential $V_\delta$ starting at $S_3$ with zero speed will reach a point $q = S_2 - \delta$, $p = -\nu$, with $\nu > 0$ as small as we wish. As soon as $\nu < \mu_0$, this means that the particle will reach $S_1$ without overshooting, showing that $c_{13}^*(\delta) \leq c$. Since this is true for all $c > c_{12}^*$, the proof of this case is complete.

**Case $c_{23}^* > c$:**
Since the solution $u$ satisfies $u(x,t) \in [S_1, S_3]$ for all $t \geq 0$, we can change $f$ for $x < S_1$ without affecting the solution $u$. We will do this in such a way as to make the critical damping from the left side less than $c_{23}^*$. Define

$$f_\lambda(x) = \begin{cases} f(x) \,, & \text{if } x \geq S_1, \\ \lambda \cdot (x - S_1) \,, & \text{if } x < S_1. \end{cases}$$

We choose $0 < \lambda < \left(c_{23}^*/2\right)^2$. Next, we fix $d$ satisfying

$$2\sqrt{\lambda} \, < \, d \, < \, c_{23}^* \,.$$

Consider now the "evolution" equation

$$\begin{aligned} q' &= p \,, \\ p' &= -d \cdot p - f_\lambda(q) \,. \end{aligned} \tag{3.3}$$

If we release a particle at $S_3$ with zero speed it will—due to the definition of $c_{23}^*$ and the choice of $d < c_{23}^*$—overshoot at $S_1$. Since the potential associated to $f_\lambda$ is increasing (to $\infty$) as $x \to -\infty$, the particle will eventually return to $S_1$ from the left. This time, it will *not cross $S_1$ again*, because we have chosen $d > 2\sqrt{\lambda}$. Thus we have constructed a front for the force $f_\lambda$, travelling with speed $d$ which connects $S_3$ to $S_1$. This front is *not monotone*, it overshoots exactly once; we have argued earlier that a monotone front does not exist in this case. Let us denote $g_{\lambda,d}(x)$ the front obtained in this way—it is unique up to translation. By choosing a suitable translation we may and will assume that $g_{\lambda,d}(x) < u(x,0)$. Applying the Maximum Principle to the problem $\dot{u} = u'' + f_\lambda(u)$, we see that

$$g_{\lambda,d}(x - td) \, < \, u(x,t) \,,$$

for all $x$ and $t$. Thus, $u(x + td', t) \geq g_{\lambda,d}(x + t \cdot (d' - d))$, for all $d'$, and hence

$$\lim_{t \to \infty} u(x + td', t) \, = \, S_3 \,,$$

when $d' < d$. Since this holds for every $d$ satisfying $2\sqrt{\lambda} < d < c_{23}^*$, the assertion follows. We have covered all cases and the proof of Proposition 3.1 is complete.

**Discussion.** We have seen that in one-component problems the situation is well-understood and that explicit conditions for the occurrence of dual fronts can be given. However, the reader should be aware that we are dealing here with a situation which is generically unstable with respect to changes in the potential. Namely, a generic change of $f$ above will destroy the critical point at $S_2$ so that it either disappears or splits into a maximum and a minimum. (This mechanism has recently been used by [PCGO] to characterize the critical speed.) Thus, the example is fully understood but not generically stable. In the next section, we consider a model where the order parameter has *two* components, and we will see that the analog of the saddle point $S_2$ remains structurally stable, so that the phenomenon becomes physically robust. On the other hand, rigorous proofs can only be given for part of the statements made, while some statements must be based on numerical experiments and a general analysis of the flow for the dynamical system.

## 4. A two-component model

To be specific, we consider the following explicit model:

$$\dot{a} = (1-\delta)a'' + \frac{\partial V}{\partial a} = (1-\delta)a'' + (1+\varepsilon)a - a^3 - \alpha ab^2 \ ,$$
$$\dot{b} = (1+\delta)b'' + \frac{\partial V}{\partial b} = (1+\delta)b'' + (1-\varepsilon)b - b^3 - \alpha a^2 b \ , \tag{4.1}$$

where

$$V(a,b) = \frac{(1+\varepsilon)}{2}a^2 + \frac{(1-\varepsilon)}{2}b^2 - \frac{1}{4}a^4 - \frac{\alpha}{2}a^2b^2 - \frac{1}{4}b^4 \ . \tag{4.2}$$

Here $\alpha > -1$, $0 < \varepsilon < 1$ and $0 < \delta < 1$ are free parameters that will be fixed later. In the discussion below, we will not use that the Eq.(4.1) derives from a potential. However, it will be intuitively slightly more appealing to have a potential $V$, so that we can think in terms of friction driven motion, as in the previous sections.

This model has four constant stationary solutions:

$$(a,b) = \begin{cases} (0,0) \equiv T_1 \ , \\ (0, \sqrt{1-\varepsilon}) \equiv T_2 \ , \\ (\sqrt{1+\varepsilon}, 0) \equiv T_3 \ , \\ (a_4, b_4) \equiv T_4 \ , \end{cases}$$

where

$$(a_4, b_4) = ((\frac{1}{1+\alpha} + \frac{\varepsilon}{1-\alpha})^{1/2}, (\frac{1}{1+\alpha} - \frac{\varepsilon}{1-\alpha})^{1/2}) \ .$$

The phenomenon of a dual front will be observed for a transition from $T_1$ to $T_3$ going through the intermediate state $T_2$. For this to happen, the unstable manifold of the state $T_2$ with respect to constant perturbations should have exactly dimension 1 in $\mathbf{R}^2$, and the final state $T_3$ should be stable. This is true if and only if

$$\frac{1-\varepsilon}{1+\varepsilon} < \alpha < \frac{1+\varepsilon}{1-\varepsilon} \ . \tag{4.3}$$

Furthermore, we need conditions which establish inequalities between the various front speeds in such a way that the phenomenon of a dual front can take place. Thus, we want $c_{23}^* < c_{13}^* < c_{12}^*$. For this to happen, we impose

$$\delta > \varepsilon \ . \tag{4.4}$$

**Conjecture 4.1.** *Under the conditions (4.3) and (4.4) one has the inequalities*

$$c_{23}^* < c_{13}^* < c_{12}^* . \tag{4.5}$$

**Some arguments in support of (4.5).** First, we can show that $c_{12}^{\min} = 2\sqrt{1-\varepsilon}\sqrt{1+\delta}$. Indeed, if we consider only orbits from $T_2$ to $T_1$ with $a \equiv 0$, then the 1-component analysis applies, and we get therefore (in the 2-component context) $c_{12}^{\min} \leq 2\sqrt{1-\varepsilon}\sqrt{1+\delta}$. This latter quantity is $2\sqrt{1+\delta}\left(\partial_b^2 V\big|_{a=b=0}\right)^{1/2}$. Because the unstable manifold at $T_2$ has dimension 1, there can be no other optimizing orbit, and the result follows. Similarly,

$$c_{13}^{\min} \leq 2\sqrt{1+\varepsilon}\sqrt{1-\delta} < c_{12}^{\min} , \tag{4.6}$$

by our choice of $\varepsilon$ and $\delta$. But now, we cannot show that the first inequality in Eq.(4.6) is an equality, because the dimension of the stable manifold at $T_3$ is 2.

Consider next a front connecting $T_2$ to $T_3$. Here, we can only give a local analysis near the point $T_2$. The flow near $T_2$ leads to a monotone function $a(x)$ and thus, we can describe the pair $\big(a(x-ct), b(x-ct)\big)$ by the graph $b(a)$ describing the front near $T_2 = (0, \sqrt{1-\varepsilon})$. We have $b(0) = (1-\varepsilon)^{1/2}$. Substituting in Eq.(4.1), we get

$$\dot{a} = (1-\delta)a'' + g_b(a) ,$$

with an effective force

$$g_b(a) = (1+\varepsilon)a - a^3 - \alpha a \left(b(a)\right)^2 .$$

We reapply the discussion of Appendix A to this force, and we get the inequalities (at least locally near $a = 0$),

$$2\sqrt{(1-\delta)g_b'(0)} \leq c_{23}^* \leq 2\sqrt{(1-\delta)\sup_{0 < a \leq a_0} g_b(a)/a} . \tag{4.7}$$

It is this argument which we cannot really make global, because of insufficient control over $b$ and its dependence on the speed $c$. On the other hand, numerical studies show that in fact $a$ is monotone and the inequalities seem to hold globally. Moreover, the selection mechanism chooses $c_{23}^* = 2\sqrt{(1-\delta)g_b'(0)}$, at least as long as the local force is concave. Thus, the situation seems to be exactly the same as the one encountered in the rigorously controlled example described in Lemma A.2.

In Fig. 6 below, we show the results of a numerical simulation. The dual front is clearly visible. On doing numerical experiments, one sees that the inequalities (4.7) hold, and that a selection mechanism takes place, which is similar to the one we rigorously established for the 1-component systems. Here, we do not have a proof. But we can illustrate what is going on in a more intuitive picture, see Fig. 7.

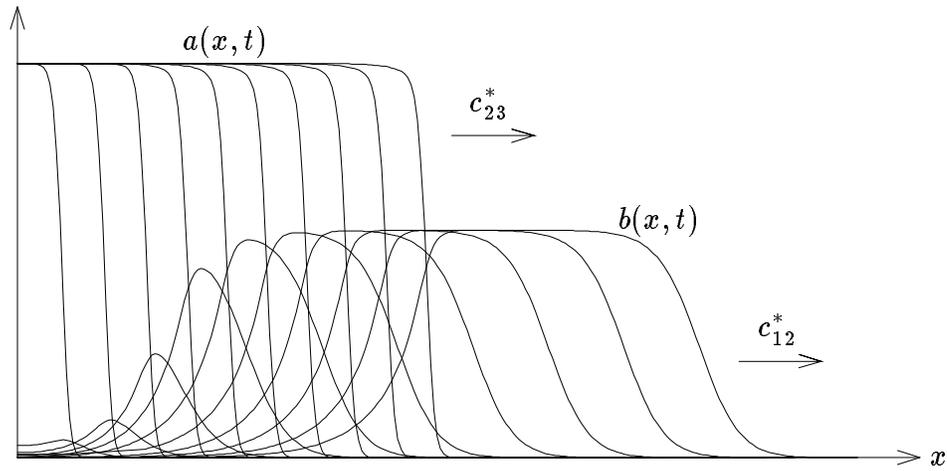

**Fig. 6**: Time evolution for the model Eq.(4.1), for the parameter values $\alpha = \varepsilon = 0.5$, $\delta = 0.8$. The time-frames shown are $t = 5$ to $t = 50$. Starting from an initial condition with very little $b$-content, a $b$-front is seen to emerge, which runs faster (with speed $c_{12}^*$) than the $a$-front (speed $c_{23}^*$). Note that the $b$ state decays to 0 with the speed $c_{23}^*$, where the $a$-amplitude grows.

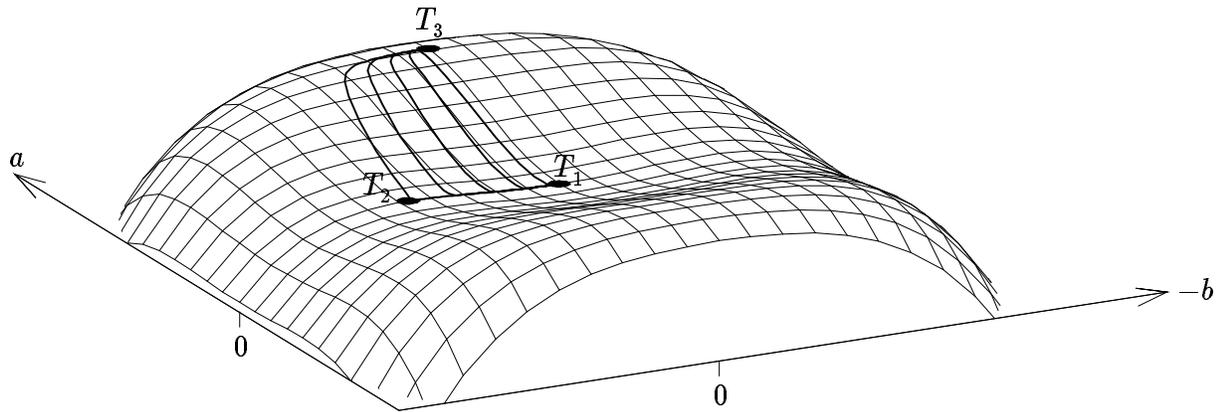

**Fig. 7**: The potential $V$ as a function of $a$ and $b$ for the same parameters as in Fig. 6. $T_3$ is a maximum, $T_1$ is a minimum, and $T_2$ is a saddle. Superimposed are orbits for times $t = 5$, $t = 12$, $t = 15$, $t = 18$ and $t = 50$. Note that, as time increases, the orbit comes closer and closer to a "circuit" $T_3$–$T_2$–$T_1$.

We are really considering a particle in the potential $V$. Since the "masses" $1 - \delta$ and $1 + \delta$ are different we can imagine for fixed $c$ the motion of a point particle in the potential $V$ with two different frictions $c/(1 - \delta) > c/(1 + \delta)$ in the $a$ and $b$ direction, respectively. The critical dampings then satisfy the inequalities (4.5).

**Discussion.** The new feature in the 2-component system is the existence of a monotone front connecting the state $T_1$ to $T_3$, whereas in the 1-component case, such fronts (connecting $S_1$ to $S_3$) did not occur. Nevertheless, the phenomenology of the 1-component systems survives, as shown by our numerical experiments, and illustrated in Fig. 6. The intuitive reason for this can be seen in Fig. 7. We see that typical initial data pick up an orbit coming closer and closer to $T_2$ as time advances, so that the transition through the intermediate state $T_2$ is "more advantageous" than a "direct transition" from $T_1$ to $T_3$. This phenomenon is *stable* under small perturbations of the equations, but harder to prove, whereas the 1-component case was unstable, but accessible to rigorous control.

## 5. Dual fronts in the magnetic model

In this section, we come back to the model of the easy-plane ferromagnet described in Eqs.(1.3) to (1.5). The discussion will now follow very closely that of the 2-component model discussed above, with a *third* "ghost front" caused by topological effects. In this model we have again three stationary solutions which will be connected by the fronts:

$$M_1 : \vartheta \, = \, \pi/2 \,, \quad \varphi' = k_0 \,,$$
$$M_2 : \vartheta \, = \, 0 \,,$$
$$M_3 : \vartheta \, = \, \pi/2 \,, \quad \varphi' = 0 \,,$$

with $k_0 > \sqrt{2}$. One can define the front speeds as before and one has

$$c_{23}^* \, = \, 2 \,, \quad c_{13}^* \, = \, 0 \,, \quad c_{12}^* \, = \, 2\sqrt{k_0^2 - 1} > 2 \,. \tag{5.1}$$

This can be seen by the following arguments:

- The front connecting $M_2$ to $M_3$ corresponds to a solution with $\varphi'(x) = 0$, and therefore the 1-component discussion of Appendix A applies. Since the function $\sin\vartheta\cos\vartheta$ is concave, this means that the speed is given by the derivative at $\vartheta = 0$, i.e., $c_{23}^* = 2$.
- The front connecting $M_1$ to $M_3$ with $\vartheta \equiv \pi/2$ has speed 0 because (1.4) reduces in this case to $\dot{k} = k''$, which is the diffusion equation.
- The speed of the front connecting $M_1$ to $M_2$ is discussed in much the same way as in the 2-component model. We assume that $\vartheta$ and $\varphi'$ are monotone functions of $x$, and consider the function $k(\vartheta) = \varphi'(\vartheta)$. Again, we have a family of concave force terms, $(1 - k(\vartheta)^2)\sin\vartheta\cos\vartheta$, and therefore the derivative at $M_1$ determines the speed, leading to $c_{12}^* = 2\sqrt{k_0^2 - 1}$, as asserted.

Applying the arguments of the preceding sections, we see the appearance of a dual front with an intermediate state which gets larger as time advances. This situation is fully confirmed by the numerical simulations of [EBS], where this phenomenon was discovered. One can see this clearly in Fig. 1 and Fig. 2, where a numerical simulation for this model is shown. Note that there is a third front appearing between the leading front travelling with speed $c_{12}^*$ and the trailing front travelling with speed $c_{23}^*$. We will call it ghost front for reasons which will become obvious shortly.

The behavior of the solution near the intermediate state $M_2$ is most efficiently analyzed in terms of the equation for $A = \sin\vartheta\,e^{i\varphi}$, cf. Eq.(1.5). Since $M_2$ corresponds to $A = 0$, we may linearize Eq.(1.5), leading to the trivial equation

$$\dot{A} = A'' + A \ .$$

The asymptotics of the trailing front leads to a factor $A_1\,e^{-x+2t}$ in the laboratory frame, while the asymptotics of the trailing part of the leading front is

$$A_2\,e^{\lambda x + (1+\lambda^2)t} \ .$$

The constant $\lambda$ can be determined from the leading part of the leading front, by observing [EBS] that the r.h.s. of Eq.(1.3) only depends on the derivative of $\varphi$ and thus leads to a conserved quantity $C$, given by

$$\partial_\xi k + c_{12}^* k + 2k\partial_\xi(\ln\sin\vartheta) = C \ ,$$

where, $\xi = x - c_{12}^* t$. Thus, the leading part of the leading front determines its trailing part, and hence $\lambda$. The explicit expression for $\lambda$ is given in [EBS].

Coming back to the problem of the ghost front, we want to determine the position of the zeros of the sum $A_1\,e^{-x+2t} + A_2\,e^{\lambda x + (1+\lambda^2)t}$, as a function of time. Solving for the zeros leads to the equation

$$(\lambda + 1)x + (1 + \lambda^2 - 2)t = \log(-A_1/A_2) + 2i\pi n \ ,$$

with $n \in \mathbf{Z}$, so that if we have one zero at $(x_0, t_0)$, then the others are—in this linear approximation—at

$$\left( x_0 + 2\pi n\frac{(\operatorname{Im}\lambda)^2 + 1 - (\operatorname{Re}\lambda)^2}{|1+\lambda|^2\operatorname{Im}\lambda} \ , \ t_0 + 2\pi n\frac{1 + \operatorname{Re}\lambda}{|1+\lambda|^2\operatorname{Im}\lambda} \right) \ .$$

Thus, the zeros seem to move with speed

$$c_{\text{ghost}} = 1 - \operatorname{Re}\lambda + \frac{(\operatorname{Im}\lambda)^2}{1 + \operatorname{Re}\lambda} \ .$$

This third speed is clearly visible in Fig. 2. The prediction for $c_{\text{ghost}}$ made in [EBS, Eq.(19)] introduced some spurious higher corrections in $\lambda$ but differs by less than 2% from the one made here. They both agree very well with the findings from the simulations.

## Conclusion

In this paper we have investigated the splitting of fronts propagating into linearly unstable states (dual fronts). Here, a single propagating front defining an interface between an unstable state and a stable final state is either unstable or may not exist. Instead, *two* fronts propagating with different velocities will emerge. They build up a linearly unstable intermediate state whose size grows linearly in time. In the frame moving with the leading (i.e., faster) front the intermediate state is only convectively unstable but it is absolutely unstable in all frames travelling slower than the trailing (i.e., slower) front.

This scenario can be rigorously proved for a structurally unstable 1-component model. Furthermore, we have good control for a generic 2-component reaction-diffusion model. Finally, we can explain in the same way this phenomenon in the magnetic model [EBS] where it was first discovered. It turns out that a *third* front can be observed in this model: The phase-slip front which is a ghost front resulting from the linear superposition of the tail of the leading front and the head of the trailing one. Such ghost fronts in the intermediate state are generally to be expected if the stable state and the unstable state are spatially periodic with different wavelengths.

The main mechanism responsible for dual fronts can be understood in the following intuitive way: Suppose we disturb locally and weakly the unstable state. In the early evolution the dynamics is governed by the linearized equation of motion. Thus the perturbation can be understood as a linear superposition of pulses, each of them governed by a linear equation of motion which decouples from all others (e.g., pulses in $a$ and $b$ in the 2-component model). A necessary condition for front splitting is the fact that the pulse which expands fastest leads to an unstable state from which another perturbation will drive the system into a stable state. This is the meaning of the inequality $c_{12}^* > c_{13}^*$.

**Acknowledgements.** This work was partially supported by the Fonds National Suisse.

## Appendix A. Definition of the critical speed

We have given an "intuitive" definition of $c_{\ell r}^*$ in Section 2. Here, we present its mathematical counterpart, as given in [AW], for $f$ as in Section 2. For monotone fronts, we can rewrite Eq.(2.3) as

$$\frac{dp}{dq} = -c - \frac{f(q)}{p} . \tag{A.1}$$

Let $p_{c,\varepsilon}(q)$ be the solution of Eq.(A.1) in the strip $(q, p) \in [\ell, r] \times (-\infty, 0]$ with initial condition $p_{c,\varepsilon}(\ell) = -\varepsilon$. If $p_{c,\varepsilon}(\tilde{q}) = 0$ for some $\tilde{q} > \ell$, then we define $p_{c,\varepsilon}(q) \equiv 0$ for all $q > \tilde{q}$. Since these curves cannot cross, it follows that they are monotone in $\varepsilon$, that is, $\varepsilon \le \varepsilon'$ implies $p_{c,\varepsilon}(q) \ge p_{c,\varepsilon'}(q)$ for all $q \in [\ell, r]$. Since $p_{c,\varepsilon}(\cdot) \le 0$, it follows that the limit

$$p_c(q) = \lim_{\varepsilon \downarrow 0} p_{c,\varepsilon}(q)$$

exists and is a curve lying in $[\ell, r] \times (-\infty, 0]$. Finally, we define

$$c_{\ell r}^* = \inf\{c > 0 \mid p_c(r) < 0\} . \tag{A.2}$$

In some special situations it is possible to determine the value of $c^*_{\ell r}$ very easily from the nonlinearity $f$, namely when

$$\sup_{r > x > \ell} \frac{f(x)}{x - \ell} = f'(\ell) \,,$$

and $f$ is positive on $(\ell, r)$. Then,

$$c^*_{\ell r} = 2\sqrt{f'(\ell)} \,.$$

A more detailed statement can be based on the following definition. For simplicity, we assume $\ell = 0$, $r = 1$. We also denote $c^* = c^*_{01}$.

**Definition.** Define $d^- = f'(0)$, $d^+ = \sup_{0 < x \leq 1} f(x)/x$, and then $c^{\pm} = 2\sqrt{d^{\pm}}$.

Then, in their fundamental work [AW], Aronson and Weinberger prove the following results:

**Proposition A.1.** *Let $f(0) = f(1) = 0$ and $f(x) > 0$ for $x \in (0, 1)$. Then*

   *i) For any $c \geq c^*$ there exists an orbit $P_c$ of Eq.(A.1) connecting $(q, p) = (0, 0)$ to $(q, p) = (1, 0)$, which stays in the strip $(q, p) \in [0, 1] \times (-\infty, 0]$.*

  *ii) No such solution exists for $c < c^*$.*

 *iii) If $c^* > 2\sqrt{f'(0)}$, then the connecting orbit $P_c$ coincides with $p_c$, if and only if $c = c^*$.*

 *iv) If $c \geq c^-$ then $p_c$ has near $(0, 0)$ the slope $\lambda_c$, where*

$$\lambda_c = -\tfrac{1}{2}\left(c + \sqrt{c^2 - 4f'(0)}\right) \,, \tag{A.3}$$

    *and every other curve through $(0, 0)$ has slope $\tilde{\lambda}_c$ with*

$$\tilde{\lambda}_c = -\tfrac{1}{2}\left(c - \sqrt{c^2 - 4f'(0)}\right) \,.$$

 *v) (Selection) The critical speed $c^*$ satisfies $c^- \leq c^* \leq c^+$. If we take a localized positive perturbation of a front as initial condition, then the solution to Eq.(2.1) will eventually travel with speed $c^*$. The shape of the selected front is given by the curve $P_{c^*}$.*

**Proof.** The proofs can be found in the following places in [AW].

   i) Theorem 4.1 and Remark 1.

  ii) Lemma 4.3.

 iii) Theorem 4.1 and Remark 1.

 iv) Proposition 4.4.

 v) Proposition 4.2, the definition of $c^*$, Eq.(A.2), and Theorem 5.1 and 5.2.

**Remarks.**

  – Since i) and ii) say that a connecting orbit exists for $c \geq c^*$, but not for $c < c^*$, this means that $c^{\min} = c^*$, as asserted in Section 2.

  – If $f'(1) \neq 0$ then $(0, 1)$ has a one-dimensional unstable manifold for the flow defined by Eq.(2.3) and the orbit $p_c$ mentioned in i) is unique.

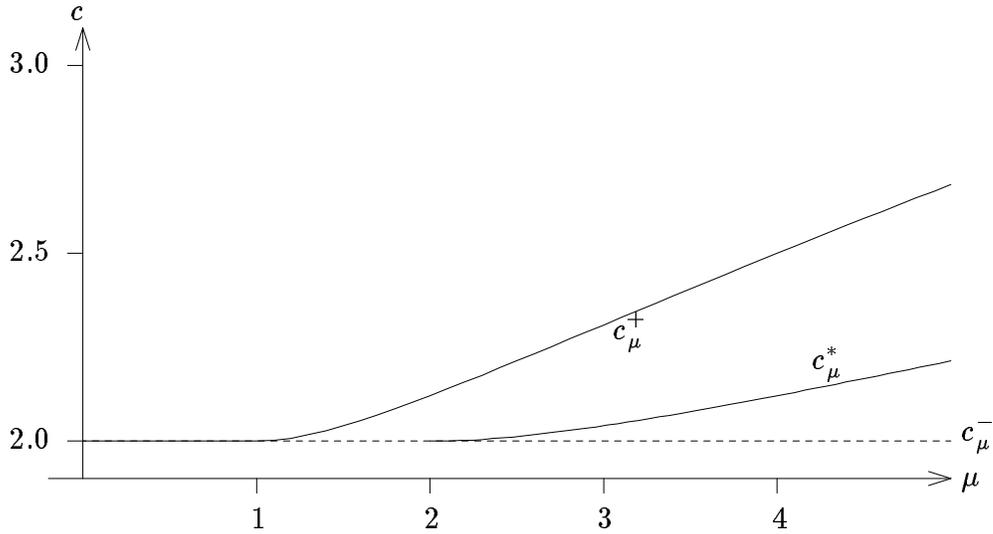

**Fig. 8**: The three curves $c_\mu^-$, $c_\mu^+$, and the front speed $c_\mu^*$ which is only different from $c_\mu^-$ for $\mu > 2$ although $c_\mu^+$ and $c_\mu^-$ already differ for $\mu > 1$.

- The result v) above has the following consequence: If $f$ is concave, then $c^* = c^- = 2\sqrt{f'(0)}$.
- One can improve the upper and lower bounds by considering polygonal approximations to $f$ instead of the linear ones used above.

### A.1. Bounds on the selected speed

This subsection illustrates the problem that the selected speed of a front is difficult to determine, and does not follow from any simple functionals of $f$, such as $f'(0)$ or $\max_{x>0} f(x)/x$. This in turn is responsible for the difficulty of giving bounds on the selected speeds for problems with more than 1 component.

The question we ask is whether it is possible that $c^- = c^*$, even when $c^- \neq c^+$. This question has been answered positively by Hadeler and Rothe [HR]. We give a slightly different account of the argument of [BBDKL].

Consider the special case of $f_\mu(u) = u + (\mu - 1)u^2 - \mu u^3$. Define $c_\mu^\pm$ in analogy with $c^\pm$, for the function $f_\mu$. We have the

**Lemma A.2.** *For $f_\mu$ as above, the critical speed $c_\mu^*$ satisfies $c_\mu^* = 2$ for $\mu \in [0, 2]$, but $c_\mu^+ > c_\mu^-$ for $\mu > 1$.*

**Remark.** The various $c$'s are shown in Fig. 8. One has

$$c_\mu^+ = \begin{cases} 2, & \text{if } \mu < 1, \\ \sqrt{\mu} + 1/\sqrt{\mu}, & \text{if } \mu \geq 1. \end{cases}$$

**Proof.** Choose the function $v(\xi) = \frac{1}{2}\left(1 - \tanh(\gamma\zeta)\right)$. Then, some algebra leads to

$$v'' + cv' + f_\mu(v) \; = \; \frac{1}{2}(1 - \tanh(\gamma\zeta))\left(2\gamma^2\tanh(\gamma\zeta) - c\gamma + \frac{1}{2} + \frac{\mu}{4}(1 - \tanh(\gamma\zeta))\right) . \; \text{(A.4)}$$

This is equal to 0 if and only if $\gamma = \pm\frac{1}{2}\sqrt{\frac{\mu}{2}}$, and

$$c \; = \; \pm\left(\sqrt{\frac{\mu}{2}} + \sqrt{\frac{2}{\mu}}\right) \; \equiv \; C_\mu \; .$$

(The signs are the same in both expressions.) Since we are interested in $c > 0$, only the plus sign is interesting for us. The calculation above means that we have found a positive front solution $v = v_\mu$, moving with speed $C_\mu$ to the right. Therefore $c_\mu^* \leq C_\mu$. But we also know from Proposition A.1, v) that $c_\mu^* \geq 2$, since $c_\mu^- = 2$ for all $\mu$. Furthermore, for $\mu \leq 1$, the function $f_\mu$ is concave in $u \in [0,1]$, and therefore $c_\mu^* = 2$ for $\mu \leq 1$. Since we have found a function $v$ for $\mu = 2$ which produces a front with speed $C_\mu = 2$, we find $c_2^* \leq 2$ and therefore, in fact $c_2^* = 2$. We shall complete the proof of Lemma A.2 by showing that $c_\mu^*$ is a monotone (increasing) function of $\mu$. Thus $c_\mu^* = 2$ for all $\mu \in [1,2]$, and we get the desired result.

To prove the monotonicity, observe first that $f_\mu(x)$ is a monotone increasing function of $\mu$, for $x \in [0,1]$. So, for all $c$, and for all $\mu \leq 2$, one has

$$p_{c,\mu}(q) \; \leq \; p_{c,\mu=2}(q) \; ,$$

for all $q \in [0,1]$. Therefore it follows from Eq.(A.2) that $c_\mu^* \leq c_2^*$, if $\mu \leq 2$.

Next we show

**Lemma A.3.** *For $f_\mu$ as above, the critical speed $c_\mu^*$ satisfies $c_\mu^* = C_\mu$ for $\mu > 2$.*

**Proof.** First we note that for $c = C_\mu$ the connecting orbit $v$ is a trajectory $p_c$ defined by the above mentioned $\varepsilon$-limit. This can be seen by using Proposition A.1, iv) and simply verifying that $v$ has near $(0,0)$ slope $\lambda_c$, see (A.3). This result implies together with Proposition A.1, iii) that $c_\mu^* = C_\mu$.

**Remark.** The proof really shows that if there is a monotone *connecting* orbit with speed $c_f$ which coincides with the "fast" stable manifold of the point $(0,0)$, then we have $c^* = c_f$.